\pdfoutput=1
\documentclass[twocolumn,prl,showpacs,aps,superscriptaddress]{revtex4-1}

\usepackage[english]{babel}
\usepackage[ansinew]{inputenc}
\usepackage{times}
\usepackage{graphicx}
\usepackage{graphics}
\usepackage{amsmath}
\usepackage{amsfonts}
\usepackage{amssymb}
\usepackage{amsbsy}
\usepackage{dsfont}
\usepackage{epstopdf}
\usepackage{makeidx}
\usepackage{subfigure}
\usepackage{color} 
\usepackage{pgf}
\usepackage{bm}
\usepackage{tikz} \usepackage[normalem]{ulem}

\newcommand{\be}{\begin{equation}}
\newcommand{\ee}{\end{equation}}
\newcommand{\bea}{\begin{eqnarray}}
\newcommand{\eea}{\end{eqnarray}}

\makeindex
\begin{document}

\title{Nonlinear Electromagnetic Interactions in Energetic Materials}

\author{M. A. Wood}
\affiliation{School of Materials Engineering and Birck Nanotechnology Center,
Purdue University, West Lafayette, Indiana, 47907, USA}
\affiliation{Theoretical Division, MS B213, Los Alamos National Laboratory, Los Alamos, New Mexico 87545, USA}
\author{D. A. R. Dalvit}
\affiliation{Theoretical Division, MS B213, Los Alamos National Laboratory, Los Alamos, New Mexico 87545, USA}
\author{D. S. Moore}
\affiliation{Explosive Science and Shock Physics Division, MS P952, Los Alamos National Laboratory Los Alamos, New Mexico 87545, USA}

\date{\today}

\begin{abstract}
We study the scattering of electromagnetic waves in anisotropic energetic materials. Nonlinear light-matter interactions in molecular crystals result in frequency-conversion and polarization changes. Applied electromagnetic fields of moderate intensity can induce these nonlinear effects without triggering chemical decomposition, offering a mechanism for non-ionizing identification of explosives. We use molecular dynamics simulations to compute such two-dimensional Raman spectra in the terahertz range for planar slabs made of PETN and ammonium nitrate. We discuss third-harmonic generation and polarization-conversion processes in such materials. These observed far-field spectral features of the reflected or transmitted light may serve as an alternative tool for stand-off explosive detection. 
\end{abstract}


\maketitle


\section{\label{sec:level1}Introduction}

Although remote sensing of energetic materials is a pressing current problem worldwide, there is a serious gap in the suite of technologies available to detect energetic materials, namely the use of non-ionizing radiation for bulk explosive detection \cite{Brown1015}. Terahertz spectroscopy has arisen as a promising tool for remote sensing of explosives because it can penetrate dry materials, is non-ionizing, is stand-off capable and, importantly, new sources of THz radiation are becoming available. Most previous studies of the interaction between THz radiation and explosives has concentrated in molecular fingerprinting of spectral responses of typical explosives in the 0.1-3.0 THz region \cite{Kemp2011}. However, since the involved transition frequencies are much smaller than the working temperatures, the resulting spectral lines are broadened and congested, making the fingerprint assignment difficult. 

Active nonlinear responses of explosives to electromagnetic radiation in the GHz-to-THz range offer an alternative potential route towards signatures for stand-off detection. As is well known, strong EM fields modify the optical properties of matter, resulting in a non-linear relation between the applied field and the electric polarizability \cite{Boyd1992},
${\bf P} = \epsilon_0 \chi^{(1)}  \cdot {\bf E} + \epsilon_0 \chi^{(2)} \cdot {\bf E} {\bf E} + \epsilon_0 \chi^{3} \cdot {\bf E} {\bf E} {\bf E} + \ldots$.  Here $\chi^{(1)}$ is the linear susceptibility tensor, $\chi^{(i>1)}$ are the non-linear susceptibility tensors, and 
$\epsilon_0$ is the permittivity of vacuum. A myriad of nonlinear effects are possible, including frequency mixing, Pockels and Kerr effects, and Raman scattering \cite{Landau1984,Jackson1998}. In addition, electromagnetic heating of energetic materials, particularly heterogeneous ones, generates the appearance of
``hot spots", regions of concentration of electric fields and, therefore, possibly enhanced nonlinear effects. These hot spots also result in locally enhanced temperatures that modify the optical properties of the explosive material by forcing atoms/electrons to explore anharmonic regions of the energy landscape resulting in further nonlinear effects. 

Recently Katz {\it et. al}  \cite{Katz2014} have employed molecular dynamics (MD) to simulate the two-dimensional spectroscopy of a few bulk explosive crystals under an applied THz linearly polarized electric field that is homogeneous throughout the bulk material. They found frequency conversion Raman-like effects both in the co- and cross-polarized optical responses, which can serve as potential alternative fingerprints of explosives. A more realistic remote sensing scenario, though, would involve an EM wave impinging onto an air-explosive interface and the evaluation of the optical non-linear signatures in the far-field at stand-off distance from the energetic material. Here, we use an MD approach to simulate the simplest air-explosive planar interface, and compute absorption spectra, frequency-conversion tensors, and far-field response for various energetic materials, including PETN and ammonium nitrate. Depending on the magnitude of the applied electromagnetic field, the incident pulse can simply force the molecular system to make nonlinear excursions without inducing chemistry (``tickling the dragon"), or trigger decomposition of the various high explosive crystals leading to deflagration and detonation \cite{MoorePatent}. In the present study we consider EM pulses strong enough to result in non-negligible nonlinearities in the optical response of the energetic materials, but not so strong as to induce chemical reactions.


\section{\label{sec:level1}Frequency Conversion in Light-Matter Interactions}

\subsection{\label{sec:level2}Scattering of electromagnetic waves in molecular crystals}

One of the most important processes in the propagation of electromagnetic waves in transparent media is scattering, in which small intensity scattered waves are produced whose wave-vectors, frequencies, and states of polarization are different from the incoming wave \cite{Landau1984,Jackson1998}. Scattering is the result of the change in the motion of the charges in the medium as a result of an incident electromagnetic field. In the following, we briefly review the classical theory of scattering. 
Let us consider a linearly polarized input plane-wave electromagnetic field of the form 
\begin{equation}
{\bf E}_{\rm in}({\bf x},t)={\bf e}_{\rm in} E_{\rm in} e^{i {\bf k}_{\rm in} \cdot {\bf x}} e^{-(t-t_0)^2/2 \sigma^2} \cos(\omega_{\rm in} t),
\label{eq1}
\end{equation}
where $E_{\rm in}$ is the amplitude of the wave, ${\bf k}_{\rm in}$ is the input wave-vector ($k_{\rm in}=|{\bf k}_{\rm in}|=\omega_{\rm in}/c$), and
${\bf e}_{\rm in}$ is the input polarization unit vector (${\bf e}_{\rm in} \cdot {\bf k}_{\rm in}=0$). The Gaussian pulse is centered at $t_0$, has a width $\sigma$ and carrier frequency $\omega_{\rm in}$. The field drives the charges in the molecular crystal according to the Lorentz force 
${\bf F}=e {\bf E} + ({\bf v}/c) \times {\bf B}$, and for sufficiently strong amplitudes the charges explore the nonlinear regions of the energy landscape. For a single accelerated (non-relativistic) charge following a trajectory ${\bf r}(t')$, the far-field radiated electric field is given by 
${\bf E}_{\rm rad}({\bf x},t)=(e/c) [ {\bf n} \times {\bf n} \times \dot{\boldsymbol\beta}/R ]_{\rm ret}$, 
where ${\bf n}$ is a unit vector in the direction of ${\bf R}(t')={\bf x}-{\bf r}(t')$, 
$\dot{\boldsymbol\beta}(t')=\ddot{\bf r}(t')/c$ 
(the dot denotes time derivative), and ``ret" means that the quantity in the square brackets is to be evaluated at the retarded time given by 
$t'+R(t')/c=t$. Note that since the observation point is assumed to be far away from the region of space where the acceleration takes place, the unit vector ${\bf n}$ 
is almost constant in time. The corresponding Fourier spectrum is given by
\begin{equation}
{\bf E}_{\rm rad}({\bf x},\omega) \!=\! \frac{e/c}{\sqrt{2 \pi}}  \int_{-\infty}^{\infty} \!\! dt' 
[{\bf n} \!\times \!{\bf n} \!\times \!\ddot{{\bf r}}(t') ] 
\frac{e^{i \omega [t'-{\bf n} \cdot {\bf r}(t')/c]}}{R(t')}.
\label{eq2}
\end{equation}
For a collection of $N$ accelerated charges, one makes the replacement 
$e [{\bf n} \!\times \!{\bf n} \!\times \!\ddot{{\bf r}}(t') ] R^{-1}(t') e^{-i(\omega/c) {\bf n} \cdot {\bf r}(t')} \rightarrow 
\sum_{j=1}^N e_j [{\bf n}_j \!\times \!{\bf n}_j \!\times \!\ddot{{\bf r}}_j(t') ]  R_j^{-1}(t') e^{-i(\omega/c) {\bf n}_j \cdot {\bf r}_j(t')}$.
Furthermore, because the observation point is very far from the accelerated charges, one can approximate
$R_j(t') \approx R_0$ and ${\bf n}_j \approx {\bf n}$, where $R_0$ is the distance between the observation point 
${\bf x}$ and an origin of coordinates within the scattering volume, and the unit vector ${\bf n}$ is along that direction. 
Note that the acceleration of each charge $\ddot{\bf r}_j$ within the molecular crystal depends on the input electromagnetic field, and in particular is a function of the input frequency $\omega_{\rm in}$. Most of the far-field radiation is emitted at the same frequency as the input one, 
$\omega=\omega_{\rm in}$ (Rayleigh scattering), but a small portion can be emitted at a frequency different from the incident one, 
$\omega \neq \omega_{\rm in}$ (Raman scattering) due to non-linear light-matter interactions within the material. A further simplification takes place when the emitted (and also the incident) light have wavelengths $\lambda=c/\omega$ much larger than the typical linear size of the scatterer. In this case $(\omega/c) {\bf n}_j \cdot {\bf r}_j \ll 1$, and one can replace the exponentials by unity, $e^{-i (\omega/c) {\bf n}_j \cdot {\bf r}_j(t')} \approx 1$; this corresponds to the leading order in a multipolar expansion. In the following we will consider THz radiation impinging on energetic materials of typical size on the order of tens of nanometers, for which this is an excellent approximation.  

The input electromagnetic field polarizes the molecular crystal, and generates electric dipole moments ${\bf p}_{\alpha}(t')$ at each atomic position ${\bf r}_{\alpha}$. Each of these dipole moments corresponds to a pair $(j,j')_{\alpha}$ of terms in the above summations over $j$, 
such that ${\bf p}_{\alpha}(t')=| e_{j\alpha}| [ {\bf r}_{j\alpha}(t')-{\bf r}_{j'\alpha}(t')]$. We can then write the following expression for the far-field Fourier spectrum of the emitted light by the molecular crystal
\begin{equation}
{\bf E}_{\rm rad}({\bf x},\omega) \!=\! \frac{1}{\sqrt{2 \pi}}  \frac{1}{c R_0} 
{\bf n} \! \times \! {\bf n} \! \times \! \int_{-\infty}^{\infty} \!\! dt'  e^{-i \omega t'} \sum_{\alpha}  \ddot{{\bf p}}_{\alpha}(t').
\label{eq3}
\end{equation} 
Therefore, the far-field frequency spectra can be obtained from the Fourier transform of the second time-derivative of the full dipole moment of the system, ${\bf P} =\sum_{\alpha} {\bf p}_{\alpha}$.

It is worth mentioning that as each atom is polarized by the impinging electromagnetic field, the induced accelerated dipole produces secondary radiation that also acts as a field source for all other dipoles. The total local field on each dipole is then the sum of the external field plus all these secondary fields, and the time evolution for the system of dipoles should be solved self-consistently. However, although MD computes the dynamics of the full system of dipoles self-consistently in terms of force fields (see below), it does not take into account the back-action of the secondary electromagnetic fields from the accelerated dipoles onto the dynamics of the dipoles themselves. In the following we will assume that these secondary fields are much weaker than the incident one, and disregard local field corrections.


\begin{figure}[t]
\includegraphics{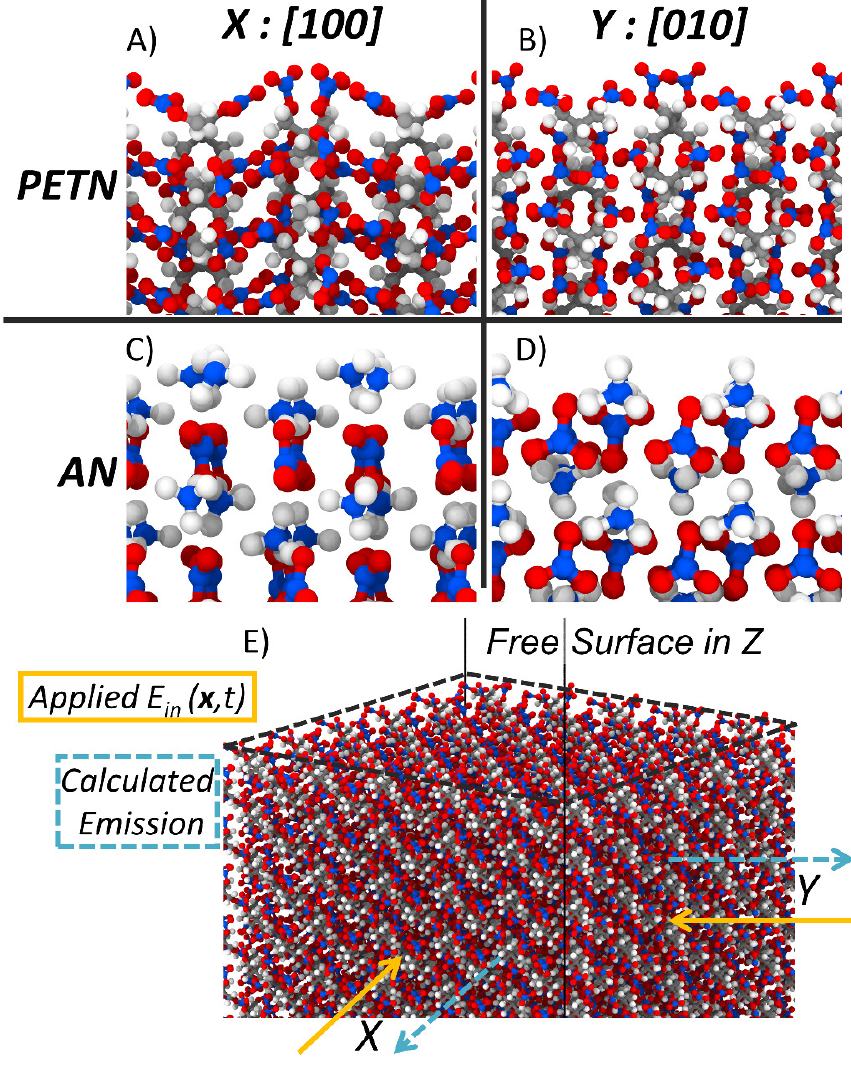}%
\caption{\label{fig:pdfart}A), C) Representative orthoscopic views of the [100] crystallographic direction for PETN and ammonium nitrate. Panels B), D) show the [010] direction in either material which exemplifies the unique molecule arrangement in either direction. E) Perspective view of the simulation domain showing how the free surface is created in MD. This simulation setuo enables realistic absorption and emission polarizations within MD. Atom colors correspond to carbon (grey), hydrogen (white), oxygen (red) and nitrogen (blue).}
\end{figure}



\begin{figure}[t]
\includegraphics{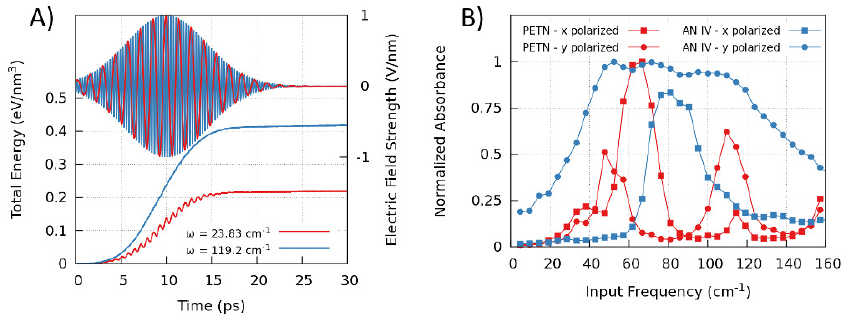}%
\caption{\label{fig:pdfart}A) Two example Gaussian electric field pulses with different carrier frequencies applied in PETN. Absorption is measured as the difference between the total energy after and before the pulse is applied. B) Absorbance (normalized by the peak value for each polarization), as a function of the input carrier frequency and input polarization, for PETN (red) and AN IV (blue). The unique molecule geometry in each direction leads to polarization-dependent absorbance.}
\end{figure}


\subsection{\label{sec:level2}Molecular dynamics methods}

Given the classical description of absorption and far field emission measurement from the previous section, we have opted to use molecular dynamics (MD) simulations as the underlying simulation tool for dynamically calculating non-linear scattering in molecular crystals. To ensure proper interaction with an externally applied electric field, a polarizable interatomic potential must be used. ReaxFF  \cite{vanDuin2001} is one such polarizable interatomic potential that is commonly used for energetic materials 
\cite{Rom2011,Zhou2012,Strachan2003}, and has been employed by several others to study the response of molecular crystals to strong electric fields \cite{Wood2014,Neyts2011,Li2015}.
Dynamic polarization is captured by allowing the partial atomic charges to be updated at each time step ($\Delta t = 0.1$ fs) with a conjugate gradient method \cite{Rappe1991} to the local minima in electrostatic energy with a tolerance set to $10^{-6}$. In addition, because the ReaxFF force field captures the reactive potential energy surface, it is intrinsically anharmonic at elevated temperatures as the system approaches transition states. It is important to note that the Gaussian electric field pulses simulated here are not sufficiently strong to initiate any chemical decomposition. However, these short electric field pulses push the system to explore anharmonic portions of the potential energy minima; this will be further discussed in Section 3. All MD simulations were run using the LAMMPS \cite{Plimpton1995} MD code with the force field parameterization contained in the Supplemental Information and described in detail in references \cite{Wood2014,Castro-Marcano2013,Wood2015}.

Recently, Katz et. al. used reactive MD to study THz emission in RDX and TATP and reported the first evidence that ReaxFF can be used to predict THz frequency conversion processes \cite{Katz2014}. In contrast to Katz {\it et. al.}, who only dealt with an infinite (bulk) material, here we consider the more physical situation of an explosive slab surrounded by air. In this case of a finite-size structure (in the $z$-direction), the use of Eqs.(\ref{eq2},\ref{eq3}) for the far-field emission is justified.  In our simulations, we will consider the faces of the slab (i.e., the free surfaces) to be orthogonal to one of the principal axis of the HE crystal (here arbitrarily chosen to be the [001] direction). Light impinges normal to the slab with polarization in the orthogonal [100] or [010] directions, and we observe emitted light in the orthogonal direction for both output polarizations. A schematic view of our simulation setup is shown in Figure 1(E). We have selected a pair of materials, pentaerythritol tetranitrate (PETN) and ammonium nitrate polymorph IV (AN IV), that exemplify a benchmark test for detection capability using THz signals \cite{McGrane2005,Holden1975}.
PETN is a molecule with both oxidizer and fuel contained on a single chemical unit, while AN is an energetic co-crystal that forms a charge neutral lattice from a basis of charged molecular species much like a ceramic material. Representative orthoscopic views of the two molecular crystals are displayed in Figure 1(A-D) \cite{Stukowski2009}.

The PETN simulation cell is generated by replicating the geometrically relaxed unit cell six times in each direction (1008 molecules, 
$7.5 \times 7.8 \times 6.3 \, {\rm nm}^3$) before removing the periodic boundary in [001] and relaxing in the isobaric-isothermal thermodynamic ensemble at 50 K and 1 atm of pressure. Similarly, the AN unit cell is replicated 12 times in each direction (729 molecule pairs, 
$4.7 \times 6.3 \times 4.8 \, {\rm nm}^3$) and relaxed using the same procedure. All simulations of an applied electric field are run in the microcanonical ensemble (constant number of particles and volume) so that any change in total energy is due to the material coupling with the electromagnetic insults. Special care must be taken when choosing the peak intensity and duration of the electric field pulse because the material needs to experience an insult strong enough to sample anharmonic regions of the potential energy surface, but remain chemically unreacted. A Gaussian pulse of width 20 ps and peak amplitude of 1 V/nm was determined to be sufficient to induce the non-linear effects of interest. 
During this pulse duration, the dipole accelerations, $\ddot{\bf p}_{\alpha}(t')$, are outputted at 4 fs intervals, and are then used to calculate the emission spectra for each carrier frequency. At each output time, the net dipole acceleration (sum on all atoms) in each Cartesian direction of the system is recorded and it is this time series that is Fourier transformed, as in Eq. (\ref{eq3}). These resultant spectra will be discussed in detail in Section 3. After this 20 ps pulse, the electric field is removed and the microcanonical dynamics are continued, which allowed for an accurate determination of the energy input.  Figure 2(A) shows a pair of example electric field pulses applied to PETN along with the total energy change due to the material coupling with this pulse. A volume normalized total energy is used here because the field is applied homogenously (long wavelength approximation) therefore making the total energy absorbed dependent on the simulation cell size. 

As a function of carrier frequency and polarization, Figure 2(B) shows the absorbance of either material normalized by the highest absorbing mode tested. Each carrier frequency is approximately $5 \, {\rm cm}^{-1}$ apart spanning the range of $0-170 \, {\rm cm}^{-1}$ in 35 steps. The strongest absorbing mode resulted in a 165 K rise for PETN and 360 K for AN; neither final temperature is expected to initiate decomposition. Interestingly, each polarization of the incident field causes a unique absorption signal due to the asymmetry in the structure along these crystallographic directions, see Figure 1. In this range of frequencies tested, most vibrations are either multi-molecular modes or small amplitude uni-molecular deformation modes. Examples of these include the nitro wag mode in PETN ($ \simeq 40 \,{\rm cm}^{-1}$) and the ammonium-umbrella mode in AN ($\simeq 280 \, {\rm cm}^{-1}$). The collection of all predicted vibrations can be obtained by Fourier transforming the time series of atomic velocities 
\cite{Berens1983}. This is known as the power spectrum, or kinetic energy per mode in the system, 
\begin{equation}
S(\omega) = \frac{\tau}{M} \sum_{j=1}^{3M} m_j \left| \sum_{n=0}^{M-1} \nu_j(n \Delta t) e^{-i 2 \pi \omega n \Delta t} \right|^2.
\label{eq4}
\end{equation}
Here, $\nu_j$ and $m_j$ are the atomic velocities and masses respectively at time $t$, $\tau$ is the sampling rate, and $M$ is the number of discrete frames to be analyzed in the Fourier transform. An overall normalization by the total kinetic energy in Eq. (\ref{eq4}) yields the vibrational density of states. The ReaxFF predicted spectra (broken down by element type) for PETN and AN at 50 K are displayed in Figure 3. 
At these low frequencies, PETN and AN exhibit mostly oxygen dominated vibrations and very few modes in PETN below $300 \, {\rm cm}^{-1}$ contain significant contribution from hydrogen, while all vibrations in AN have some hydrogen character. While not all of these modes carry a dipole moment, it is possible that the strong electric field will polarize the sample and make for uniquely radiated signals. The spectra shown in Figure 3 provide a baseline for what the frequency conversion tensor may entail. While the input pulses are able to strongly polarize the samples, we do not expect that new modes of vibration will be observed due to the lack of conformation or chemical change.

\begin{figure}[t]
\includegraphics{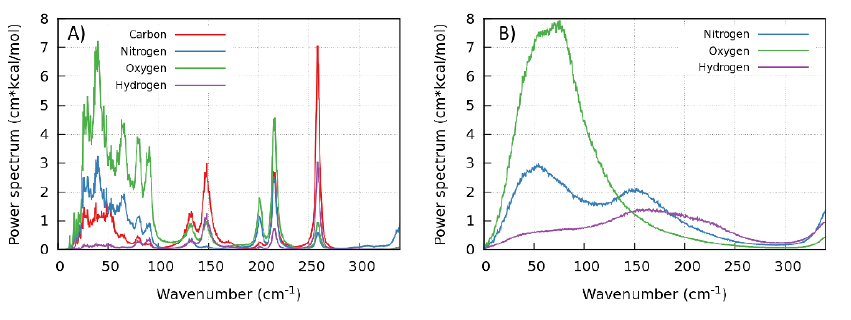}%
\caption{\label{fig:pdfart}ReaxFF calculated power spectrum for A) PETN and B) AN at 50 K and 1 atm of pressure. These spectra aid in analyzing the radiated electric field signals shown in Section 3. The sum of all element contributions normalized by total kinetic energy yields the vibrational density of states.}
\end{figure}

\begin{figure}[]
\includegraphics{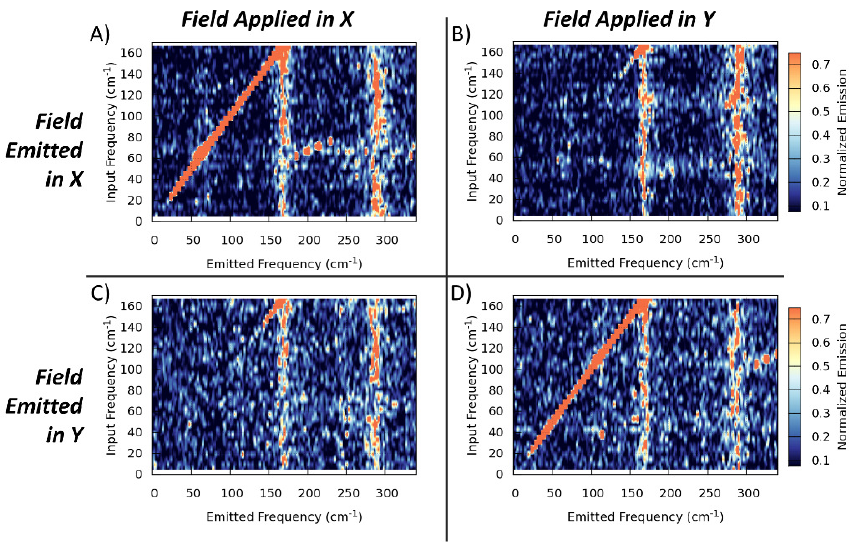}%
\caption{\label{fig:pdfart}Emission signals from PETN along the [100] ($x$ direction) for electric field pulses applied in the A) [100] and B) [010] directions. Complementary emission signals with pulses along the [010] ($y$ direction) for the C) [100] and D) [010] directions in PETN. Where the emission signals are aligned with the pulse polarization, Rayleigh scattering dominates the observed emission. Perpendicular directions do show emission at unique frequencies different from the applied pulse, confirming the frequency conversion due to internal scattering.}
\end{figure}
\begin{figure}[]
\includegraphics{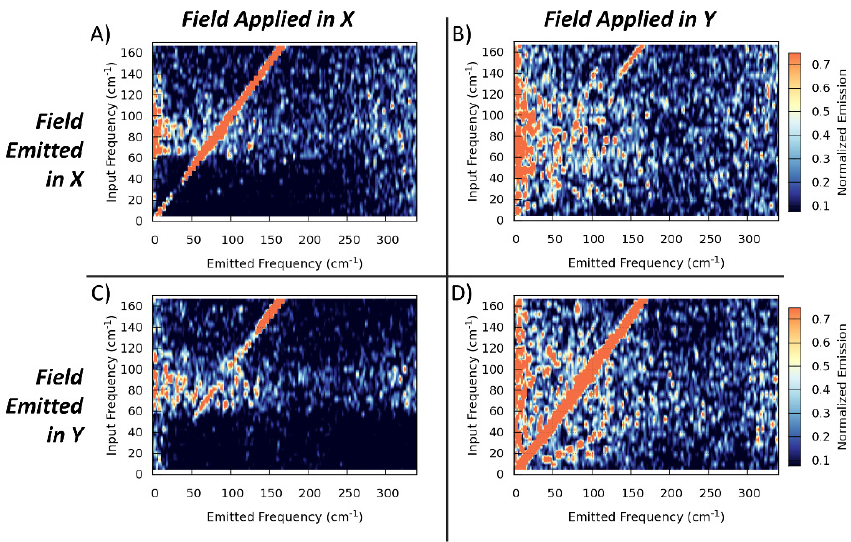}%
\caption{\label{fig:pdfart}Emission signals from AN along the [100] ($x$ direction) for electric field pulses applied in the A) [100] and B) [010] directions. Complementary emission signals with pulses along the [010] ($y$ direction) for the C) [100] and D) [010] directions in AN. Third harmonic emission is clearly seen for parallel emission directions to the applied field and is much weaker in orthogonal directions. }
\end{figure}

\section{\label{sec:level1}Results}
\begin{figure*}[t]
\includegraphics{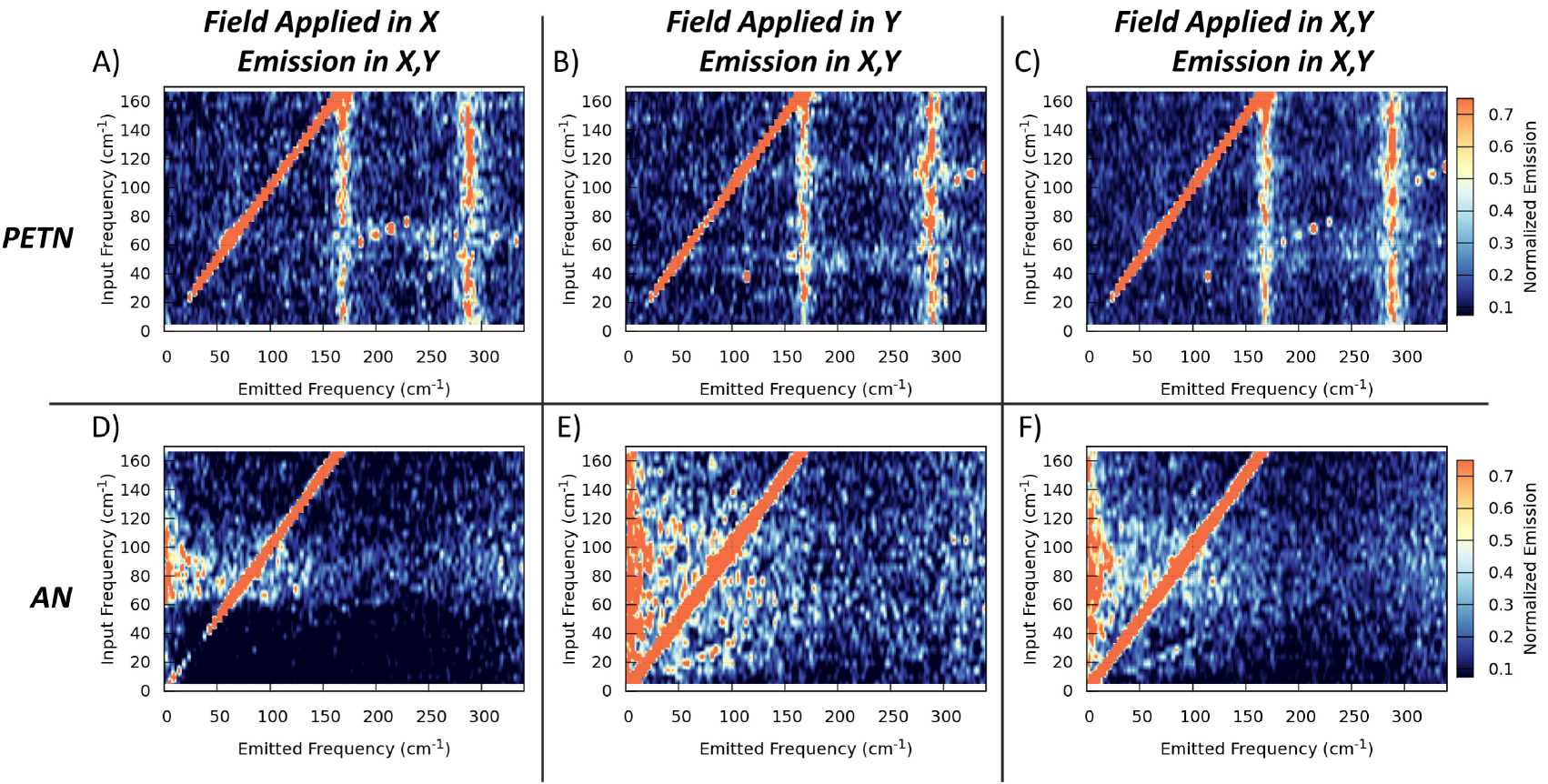}%
\caption{\label{fig:wide}Average emission signals for all output polarizations for insults applied in the [100] direction for A) PETN and D) AN along with field pulses in the orthogonal, [010] direction in panels B) and E).  Total emission signals for all absorption and exit polarizations for C) PETN and F) AN.}
\end{figure*}

In section 2.1, Eq.(\ref{eq2}) is explicitly a function of the carrier frequency of the applied Gaussian pulse; it is the goal of this work to identify unique 
${\bf E}_{\rm rad}$ traces for a range of THz carrier frequencies. For each polarization direction shown in Figure 1(E), two emission spectra are calculated from each simulation of a given carrier frequency. The collection of all emission spectra over the range of input frequencies results in a two-dimensional map of the frequency conversion for either material. These emission signals are normalized to the amount of energy absorbed; the color axis in Figures 4-6 represents the percent total emission at the given input and emitted frequency pair. Since the materials of this study are anisotropic, we expect polarization mixing even for the case of normal incidence.

The set of calculated frequency conversion maps for PETN is contained in Figure 4. Emission parallel to the applied field, shown in panels (A) and (D) of Figure 4, display distinct Rayleigh emission above an input frequency of $20 \, {\rm cm}^{-1}$. In addition, the third harmonic generation is clearly seen in Figure 4(A),(D) where the input frequency is one-third that of an existing mode in the material, see Figure 3(A). The second harmonic is forbidden since PETN has inversion symmetry \cite{Boyd1992}. Orthogonal to the applied field polarization, the emitted signal is not as well defined as the parallel emission, though unique islands of frequency conversion can be identified. In panels (B) and (C) of Figure 4, the Rayleigh emission is not uniformly seen for all carrier frequencies and varying emitted intensities are observed near $160 \, {\rm cm}^{-1}$ and $270 \, {\rm cm}^{-1}$ 
across all input frequencies.

The energetic co-crystal AN has a unique set of frequency conversion maps, as shown in Figure 5. Parallel to the applied field, the third harmonic emission is only weakly seen for a field pulse applied in the $x$-direction, Figure 5(A), but is visible for all carrier frequencies tested here. Conversely, the parallel emission for a $y$-polarized field only results in a clear third harmonic emission below $125 \, {\rm cm}^{-1}$. Orthogonal emission to the applied field, Figure 5(B) and (C), has a common Rayleigh emission at input frequencies higher than $140 \, {\rm cm}^{-1}$ but each polarization yields contrasting off-diagonal elements. Namely, the $x$-polarized pulse only shows non-linear elements for input frequencies in the range of $60-110 \, {\rm cm}^{-1}$ where modes below $25 \, {\rm cm}^{-1}$ or above $300 \, {\rm cm}^{-1}$ become strongly emitting signals.
Gathering each of these frequency conversions tensors, one can predict the far-field observed emission spectra. The rationale behind showing all polarization components in Figures 4 and 5 was to confirm that the input light is being scattered within the crystal resulting in both a polarization and frequency change. 

For the usual experimental case where the input field is polarized but the detection is not polarization sensitive, it makes sense to perform an average over the x and y polarizations for a given linearly polarized input field. The frequency conversion maps of these cases are shown in Figure 6(A) and (B) for PETN and Figure 6(D) and (E) for AN. Furthermore, in the case that the input field is unpolarized and the far-field detector is also not polarization-sensitive, it makes sense to perform an average over the x and y input polarizations, and the x and y output polarizations. The results of this computation for PETN and AN are shown in Figure 6(C) and (F), respectively. Focusing on the pair of spectra shown in Figure 6(C) and (F) these energetic materials have to be distinguished by the off-diagonal portions of the emitted light because both show strong Rayleigh signals. For PETN, there are only a handful of third harmonic frequencies that show up clearly and all of these are above $100 \, {\rm cm}^{-1}$.  These unique non-linear signals are the result of the third harmonic of the input wave coinciding or in close proximity with a peak in the vibrational spectra shown in Figure 3(A). In contrast, AN only shows a weak third harmonic generation below $100 \, {\rm cm}^{-1}$ and unique non-linear features occurring in the low frequency region below $25 \, {\rm cm}^{-1}$ for input frequencies between 40 and $120 \, {\rm cm}^{-1}$. These low frequency non-linear signals are inferred to be molecular rotations or multi-molecular modes that are activated due to strong absorption in the [010] polarized light. These unique features in the unpolarized frequency conversion maps provide signatures that an uninformed observer could use to distinguish between these two materials.

\section{\label{sec:level1}Conclusions}
In this contribution we have leveraged molecular dynamics simulations and extensions of classical electrodynamics to predict non-linear THz emission due to vibrational scattering in a pair of molecular crystals. By constraining the simulation geometries and polarizations of the external electric our predictions are easily extended to the experimentally relevant far-field measurement. In addition, by using reactive MD simulations we have avoided any approximations due to harmonic absorption or emission as well as normal mode descriptions of the molecular vibrations. While the use of a force field is an approximation of the actual physics of the problem, it does however allow for tractable simulations that provide useful and rapid 2D spectra predictions. We have shown the 2D spectra of PETN and ammonium nitrate polymorph IV that differ significantly in the THz frequency range due to changes in the location and intensity of third-harmonic emission of the input light as well as other non-linear frequency emissions. These output signals are dramatically different in both polarization and frequency from the input electric field pulse, which provides a potential for a THz detection technique with chemical specificity. With increasing pulse intensities one can imagine using the resultant chemistry from the THz pulse as an additional means for stand-off detection, but this was beyond the scope of this contribution. There would be a necessary tradeoff of the specificity of the predicted non-linear signals for chemical species detection due to the high temperature broadening of these low frequency modes of interest. The pulse intensity dependence of non-linear signals in these anisotropic energetic materials will be a subject of further study.

\section{\label{sec:level1}Acknowledgments}
\begin{acknowledgments}
We would like to thank Wilton Kort-Kamp for discussions. MAW recognizes partial support for this research by the U.S. Office of Naval Research through grant number N00014-11-0466. DARD and DSM gratefully acknowledge funding from the US DOE through the LANL LDRD program. 
\end{acknowledgments}

\end{document}